# Electric-field Control of Giant Ferronics

Baolong Zhang[1,2†], Ruihuan Duan[3†], Sobhan Subhra Mishra[1,2], Sambhu Jana[1,2], Jonghyeon Kim[4], Thomas Tan Caiwei[1,2], Yi Ji Tan[1,2], Wenhao Wang[1,2], Pang Teng Chen Ietro[1,2], Zheng Liu[3*], Ranjan Singh[5*]

[1] Division of Physics and Applied Physics, School of Physical and Mathematical Sciences, Nanyang Technological University, Singapore, Singapore.

[2] Centre for Disruptive Photonic Technologies, The Photonics Institute, Nanyang Technological University, Singapore, Singapore.

[3] School of Material Science and Engineering, Nanyang Technological University, 639798, Singapore.

[4] School of Electrical and Electronic Engineering, Nanyang Technological University, Singapore, Singapore.

[5] Department of Electrical Engineering, University of Notre Dame, Notre Dame, IN, USA.

†These authors contributed equally to this work.

*Corresponding author. Email: z.liu@ntu.edu.sg, rsingh3@nd.edu

**Ferrons are quantum excitations of electric polarization in ferroelectrics and electric analogues of magnons but have lacked direct experimental verification at room temperature. We harness the coupling of soft phonons and ferroelectric order in layered $NbOX_2$ (X = I, Br, Cl) to generate, detect, and control giant ferrons, creating a new class of ultralow-power, chip-scale terahertz (THz) sources. Multiple ferron modes produce intense, narrowband THz emission with quality factors up to 228 and radiation efficiencies up to five orders of magnitude greater than state of the art semiconductor emitters. Resonant excitation of a high-$Q$ ferron mode achieves efficiencies two orders of magnitude higher than intense lithium niobate THz sources. We further demonstrate direct, non-volatile electric-field control of ferron oscillations. These findings provide evidence for multiple ferrons and establish Ferronics as a foundational platform for light- and field-driven control of quantum order, with broad impact on ultrafast electronics, photonics, quantum technologies, and next-generation wireless communication.**

**Introduction:**

The collective excitation of ferroelectric order gives rise to a theoretically predicted quasiparticle known as the "ferron" [1,2] analogous to magnons, which represent the collective excitation of spin waves in magnetic systems.[3] While magnons have been extensively studied and directly observed in a variety of materials, the existence of ferrons has remained experimentally unverified. Previous studies have provided indirect evidence for ferrons by demonstrating electric-field-dependent transport behaviors in ferroelectric materials.[4] However, direct observation that reveals the microscopic excitation dynamics and behavior of the ferroelectric order is still lacking. The excitation of ferroelectric order involves perturbations of spontaneous polarization, which corresponds to oscillating electric dipoles, for example, a ferroelectric-order-coupled phonon mode. Similar to magnetic dipoles, the magnon, these oscillating electric dipoles are expected to emit electromagnetic radiation into free space, typically in the gigahertz to terahertz (THz) frequency range.[5] Based on this principle, we aim to provide definitive evidence for ferron excitation at room temperature by detecting coherent THz radiation emitted from ferroelectric materials following ultrafast optical excitation.

Recently, a new class of nonlinear materials has emerged: van der Waals ferroelectric materials, specifically $NbOX_2$ (X = I, Br, Cl). These materials exhibit exceptional properties, including a second-order nonlinear coefficient more than an order of magnitude higher than that of conventional nonlinear materials such as lithium niobate crystal ($LiNbO_3$) and transition metal sulfides and selenides[6–8]. This exceptional nonlinearity has been demonstrated in high-efficiency second-harmonic generation. In addition, they display weak interlayer coupling and strong chirality, which have garnered significant attention in several studies[6,9]. These remarkable properties also position these materials as promising for strong light-matter interaction and efficient excitation of the ferron mode. Furthermore, the ferroelectric and piezoelectric characteristics suggest the possibility of strain and electric-field tunable radiation[8,10].

For THz generation, most known emission mechanisms originate from transient electronic currents. In contrast, contributions from ions and lattice dynamics are typically negligible. Lattice vibrations, when present, usually manifest as absorption features rather than emission sources in THz spectra, as observed in materials such as ZnTe[11]. A few materials, such as PbTe[12], and hybrid perovskites ($CH_3NH_3PbI_3$)[13], show short-lived phonon oscillations at room temperature, resulting in weak emission peaks or strong emission peaks at extremely low temperature using molecular crystal.[14] However, coherent THz emission from ferrons at room temperature could serve as a new radiation source, offering a novel approach for controlling the radiation through the manipulation of the ferroelectric order through various approaches, including strain engineering[15] or electric field[4] which makes it promising for developing novel radiation sources.

We report highly efficient narrowband THz emission at room temperature from van der Waals ferroelectrics $NbOX_2$. Supported by Raman spectroscopy and prior calculations, we attribute this emission to multiple phonon modes coupled to ferroelectric order, constituting ferronic radiation, which is further corroborated by an electric-field-driven ferron phase-reversal.

**Results**

**Material structure and experimental layout.**

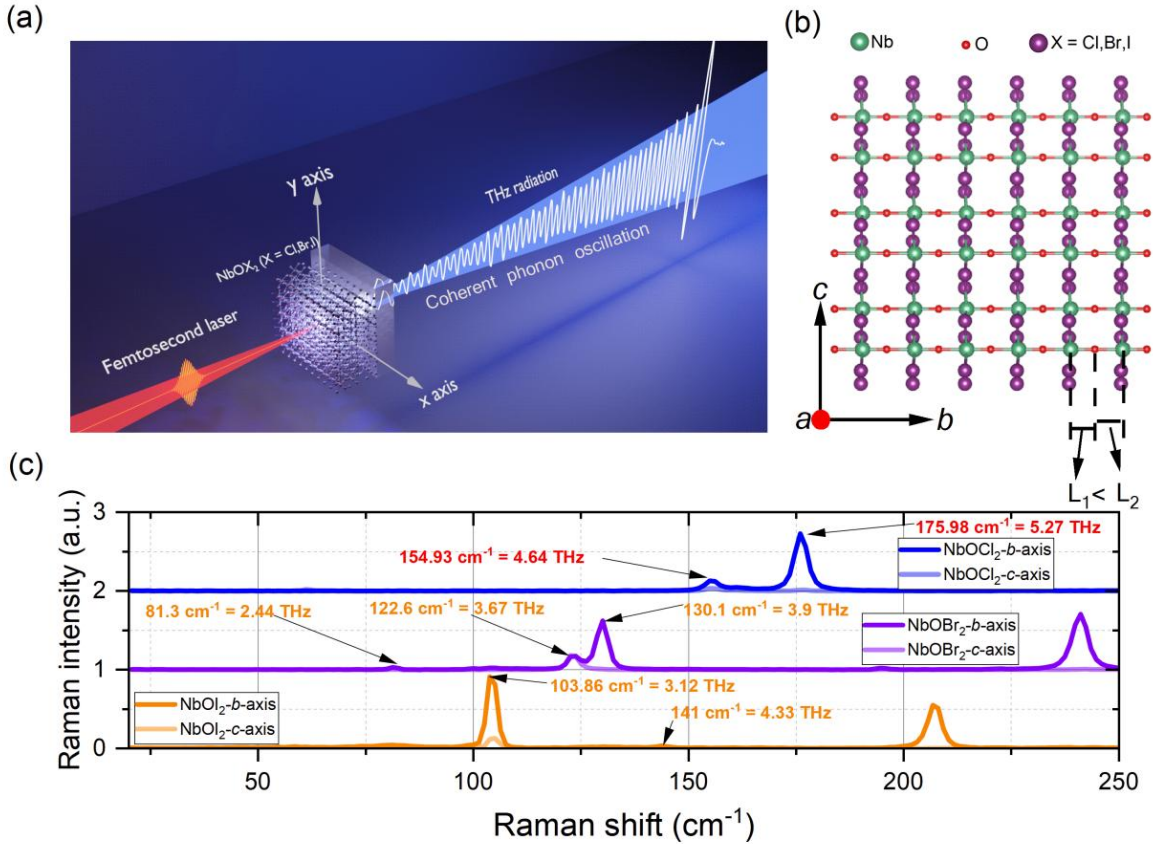

**Fig. 1. Material structure and experimental layout.** (**a**) Schematic illustration of the THz emission from NbOX$_2$ (X = I, Br, Cl). (**b**) Crystal structure of NbOX$_2$; the polar direction is along the *b*-axis. (**c**) Raman spectroscopy of NbOX$_2$ detected along *b*-axis and *c*-axis. Raman peaks (below 200 cm$^{-1}$) located at 103.86 cm$^{-1}$ and 141 cm$^{-1}$ for NbOI$_2$; 81.3 cm$^{-1}$, 122.6 cm$^{-1}$ and 130.1 cm$^{-1}$ for NbOBr$_2$; 154.93 cm$^{-1}$ and 175.98 cm$^{-1}$ for NbOCl$_2$.

Niobium oxide dihalides (NbOX$_2$) belong to a group of transition metal oxide halide compounds where the bulk NbOX$_2$ is the typical type of van der Waals layered materials constructed by stacking of NbOX$_2$ monolayers along the out-of-plane direction (*a*-axis). Each monolayer is composed of NbO$_2$X$_4$ octahedra, building up a 2D structural network through extensive interconnecting X-X edges along one planar direction (crystallographic *c*-axis) and corner-sharing O atoms along the other (*b*-axis), as shown in Fig. 1(b). Spontaneous polarization in these samples arises from the displacement of Nb atoms from the center of the NbO$_2$X$_4$ octahedra toward one of the bridging oxygen atoms. This displacement results in two distinct Nb-O bond lengths (L$_1$ < L$_2$), with the Nb-O bond oriented along the *b*-axis, which serves as the polar axis, as depicted in Fig. 1(b). In contrast, along the *c*-axis, there are Nb-Nb bonds of varying lengths but no spontaneous polarization. This contrast between polar and non-polar directions within the *b*-*c* plane gives rise to significant anisotropy in the material[6].

Fig. 1(c) shows the Raman spectra of these materials along two axes. This figure clearly shows the positions of the Raman-active phonons for each sample, such as 103.8 cm$^{-1}$, 141.0 cm$^{-1}$ for NbOI$_2$, 81.3 cm$^{-1}$, 122.6 cm$^{-1}$ and 130.1 cm$^{-1}$ for NbOBr$_2$, 154.93 cm$^{-1}$ and 175.98 cm$^{-1}$ for NbOCl$_2$. Due to the anisotropy of the material, the Raman peaks exhibit stronger signals in the polar direction compared to the non-polar axis, which is consistent with previous findings.[6–8]

The THz emission experiment utilizes a conventional pump-probe setup. NbOX$_2$ samples are first exfoliated into thin films and then transferred onto a sapphire substrate via a dry transfer method using polydimethylsiloxane (PDMS) tape. The samples are irradiated under normal incidence by 800 nm femtosecond laser, and the generated THz waves are collected and detected, as illustrated in Fig. 1(a). For simplicity, a laboratory coordinate system is used, with the horizontal direction defined as the x-axis and the vertical direction as the y-axis.

**THz ferron radiation and its anisotropic nature.**

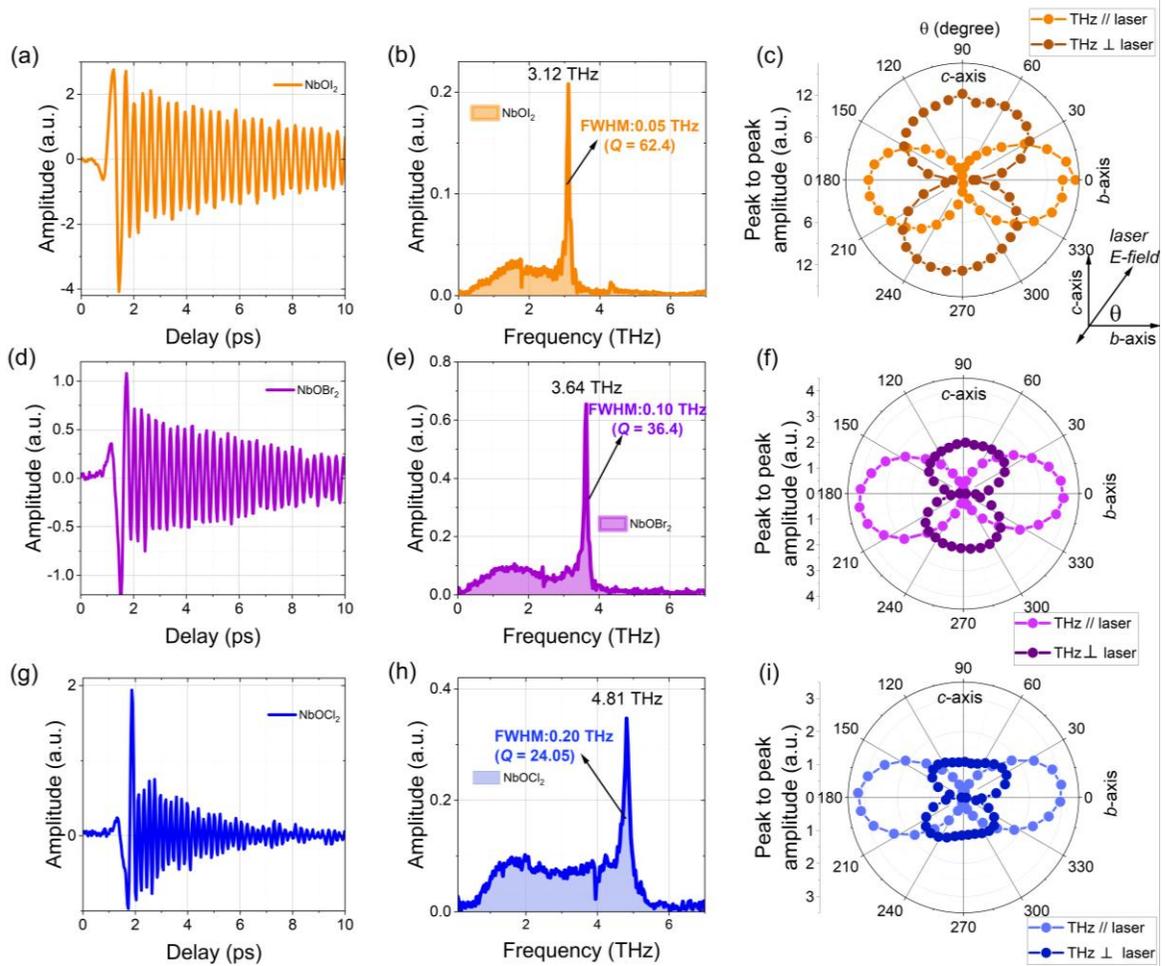

**Fig. 2. THz high quality-factor ($Q$) ferron radiation and its anisotropic nature. (a, d, g)** Time-domain THz coherent ferron oscillations radiated by NbOI$_2$ (1.62 $\mu m$), NbOBr$_2$ (1.60 $\mu m$), NbOCl$_2$ (1.12 $\mu m$) thin films. **(b, e, h)** Fast Fourier transform spectrum of THz emission from NbOI$_2$, NbOBr$_2$, NbOCl$_2$, exhibiting three distinct peaks at 3.12 THz, 3.64 THz, and 4.81 THz, respectively. The full width at half maximum (FWHM) and quality factor ($Q$) of the pronounced ferron mode is also marked in these figures. **(c, f, i)** THz emission anisotropy of NbOI$_2$, NbOBr$_2$, and NbOCl$_2$ along two directions, one where the THz polarization is parallel to the pump laser polarization, and the other where it is perpendicular to the pump laser polarization. The inset figure shows the relationship between laser polarization and THz polarization, where θ is the angle between them.

Experimentally measured THz emission data for these three materials are presented in Fig. 2, including THz time-domain waveforms, Fourier transform (FFT) spectra, as well as emission anisotropy. The THz waveform exhibits a single-cycle oscillation followed by periodic oscillations, as shown in Fig. 2(a, d, g). Notably, these materials feature high amplitudes and long lifetimes of subsequent oscillations, which appear in the spectrum as a distinct, intense emission peak superimposed on a weak and broad spectral background, as shown in Fig. 2(b, e, h). These peak positions, located at 3.12 THz (103.86 cm$^{-1}$) for NbOI$_2$, 3.64 THz (121.65 cm$^{-1}$) for NbOBr$_2$, and 4.81 THz (160.32 cm$^{-1}$) for NbOCl$_2$, agree well with the Raman spectroscopy results (as show in Fig. 1(c)), which clearly indicates that the peaks in the THz emission spectra originate from coherent phonon oscillations of each material. Specifically, previous research has reported that the oscillation of phonon in NbOI$_2$ at 3.12 THz comes from the in-phase oscillations of Nb and O atoms, which modulate the local spontaneous polarization.[16] This phonon mode coupled with ferroelectric order is a soft-phonon mode or a "ferron".[1,4] Therefore, the emitted radiation observed in this study can be referred to as coherent ferron radiation. In addition, the spectrum of the NbOI$_2$ sample exhibits three distinct emission peaks shows in Fig. 2(b), all of which display emission characteristics rather than absorption. These features can be consistently attributed to multiple ferron modes. Finally, the increase in peak frequency for each sample likely results from the decreasing mass of the halide atoms (I, Br, Cl), when considering the ferron oscillation as a classical harmonic oscillator.[17]

To accurately determine the positions of the trailing ferron frequencies, the first main THz pulse was truncated in the time domain before performing the FFT. This procedure allowed us to identify the complete spectral location of the subsequent oscillations. The unique characteristics of each ferron result in their manifestation in the emission spectrum as either distinct emission peaks or Fano resonances. For instance, the ferron frequency at 1.78 THz (see Fig. 2(b)) exhibits a Fano resonance in the radiation spectrum of the NbOI$_2$ sample, originating from the interference between broadband radiation and narrowband ferron modes, whereas other ferrons do not show this feature as their frequency either partially overlap or do not overlap with the broadband radiation.

To assess the anisotropy of THz generation in these materials, we perform azimuth-dependent measurements. Specifically, we measure the THz emission components parallel and perpendicular to the pump laser polarization while rotating the sample. The azimuth-dependent peak-to-peak amplitudes of these THz emissions, as shown in Fig. 2(c, f, i), clearly demonstrate the *C*2 symmetry of these materials. In addition, the polarity of the THz waveform reverses upon rotating the sample by 180 degrees.

**Properties of emitted giant ferronic THz radiation**

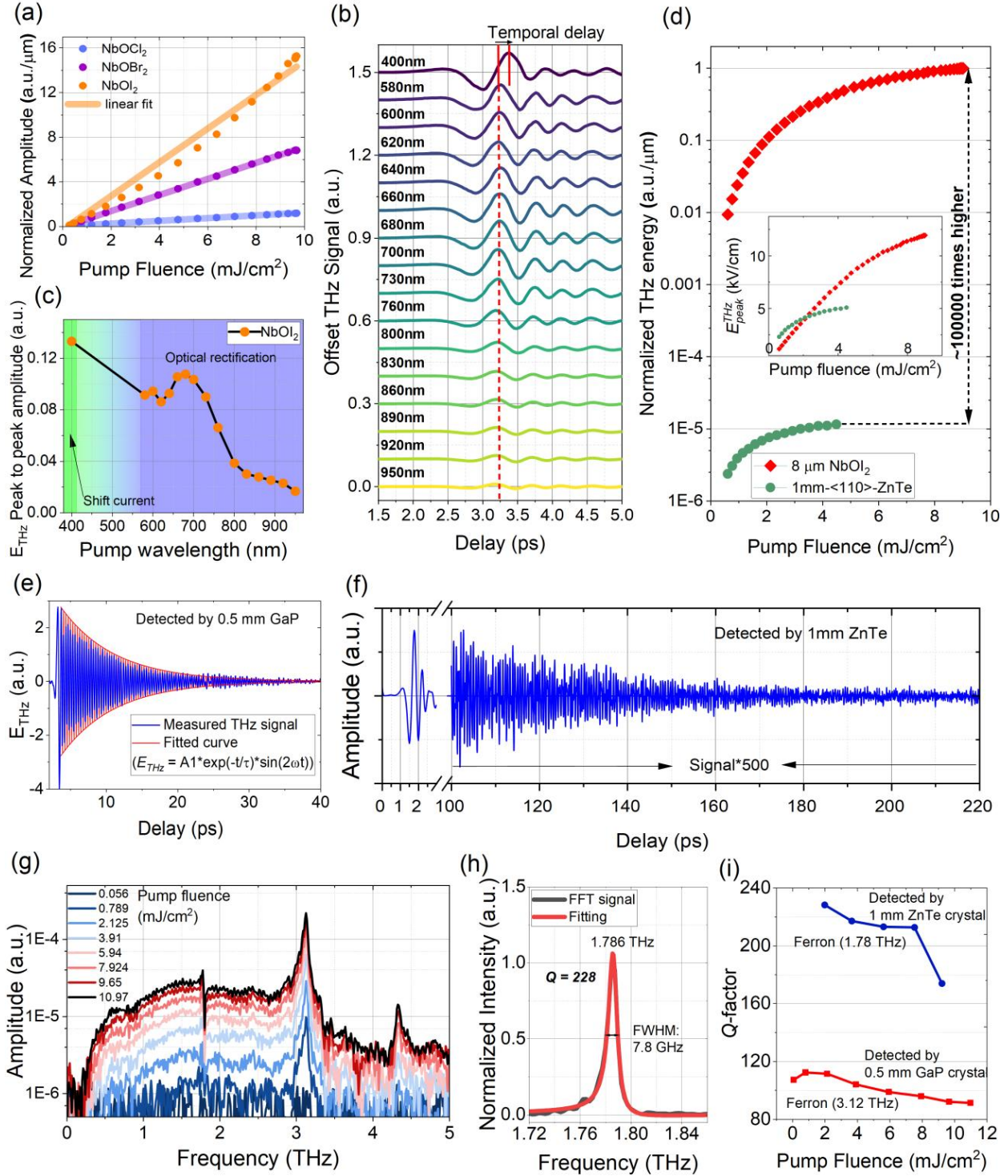

**Fig. 3. Properties of emitted ferronic THz radiation. (a)** Comparison of normalized THz emission from NbOI$_2$ (1.62 μm), NbOBr$_2$ (1.60 μm), and NbOCl$_2$ (1.12 μm) thin films. **(b)** THz time-domain waveform as a function of pump wavelength, where a clear time delay is observed at 400 nm pump and beyond 580 nm. **(c)** Peak-to-peak THz amplitude of NbOI$_2$ as a function of pump wavelength. The green region indicates shift-current dominance, the blue region indicates optical rectification dominance, and the mechanism in between is unclear. **(d)** Comparison of normalized THz energy as a function of pump fluence between an 8-μm-thick NbOI$_2$ film and a 1-mm ⟨110⟩-oriented ZnTe crystal, pumped by an 800-

nm laser and detected using 500-$\mu m$ GaP. The inset shows THz peak electric field measurement based on electro-optical effect as a function of pump fluence. **(e)** Full scan of time-domain waveform of NbOI$_2$ under 800-nm pump detected by 500-$\mu m$ GaP, the red curve represents a single-exponential decay fit with a time constant of $\tau$ =7.5 ps. **(f)** Time-domain signal from 8-μm-thick NbOI$_2$ film detected by 1-mm-thick ZnTe crystal with scan window up to 220 ps. **(g)** Corresponding FFT spectra at different pump fluences. **(h)** Example of Q-factor fitting at a pump fluence of 2 mJ/cm$^2$. **(i)** Fitted quality factor (Q-factor) at different pump fluences. For the ferron mode at 3.12 THz, we used the signal detected by 500 $\mu m$ GaP, for the ferron mode at 1.78 THz, we used the signal detected by 1 mm ZnTe crystal.

To determine the mechanism of the ferronic THz emission, we conduct pump-fluence-dependent experiments, as shown in Fig. 3(a). All three materials exhibit a strong linear dependence on pump fluence, confirming that the THz generation follows a second-order nonlinear process. Additionally, these results indicate that the NbOI$_2$ sample exhibits the strongest emission capability, followed by NbOBr$_2$ and NbOCl$_2$. Therefore, we select the NbOI$_2$ sample as the primary focus of our study.

The THz generation mechanisms can be attributed to OR and the ultrafast shift current process (as indicated by the formula below), while the photo-Dember effect and photon-drag effect are excluded, since the measurement was conducted under normal incidence[18].

$$E_{THz} \propto \frac{\partial^2 P_{OR}}{\partial t^2} + \frac{\partial J_{shift}}{\partial t}. \qquad (1)$$

Where $P_{OR}$ represents the total polarization caused by OR, and $J_{shift}$ is the shift current density. Both the OR and the ultrafast shift current mechanisms are second-order nonlinear processes, exhibiting similar trends in their dependence on pump fluence and anisotropy. The primary difference between OR and ultrafast shift current mechanism lies in whether the pump photon energy is below or above the bandgap. However, reported bandgap values in the literature vary significantly.[6,8,9,19–21]

In addition to directly comparing the bandgap with the pump photon energy, previous studies have identified a key characteristic differentiating above-bandgap from below-bandgap optical excitation: when transient optical excitation shifts from below to above the bandgap, the generated THz waveform experiences a temporal delay of approximately one-quarter cycle[22]. This delay arises from different underlying radiation mechanisms. Specifically, OR is described by $P_{OR} = \epsilon_0 \chi^{(2)}(-\omega_D; \omega_1; \omega_2) E^*(\omega_1) E^*(\omega_2)$, where $\chi^{(2)}$ is the second-order optical susceptibility and $\epsilon_0$ is the vacuum permittivity. Consequently, $E_{THz} \propto \frac{\partial^2 P_{OR}}{\partial t^2} \propto (i\omega)^2 * P_{OR}$. In contrast, the shift current mechanism is governed by $J_{shift} = \sigma^{(2)}(-\omega_D; \omega_1; \omega_2) E^*(\omega_1) E^*(\omega_2)$, where $\sigma^{(2)}$ is the second-order shift current tensor, yielding $E_{THz} \propto \frac{\partial J_{shift}}{\partial t} \propto i\omega * J_{shift}$. A comparison of these expressions reveals that the shift-current mechanism includes an additional imaginary factor compared to OR, leading to a phase delay of one quarter of a cycle (since $i = e^{i*\pi/2}$). To identify the dominant mechanism, we employed an optical parametric amplifier to vary the pump wavelength while keeping the pump fluence constant (~1 mJ/cm$^2$) and then measured the corresponding THz emission (see Fig. 3(b) and 3(c)). We observed that in the range of 950 nm to 580 nm, the THz amplitude steadily increases with no detectable temporal delay. However, at 400 nm, a clear phase delay was observed (approximately 0.167 ps, which corresponds to a quarter period of the central frequency of emitted THz pulse (1.5 THz)). This indicates that excitations from 950 nm to 580 nm are below the bandgap (dominated by OR), whereas 400 nm lies above the bandgap (dominated by ultrafast shift currents). Furthermore, the pump wavelength dependence of the THz emission (Fig. 3(c)) reflects how the second-order nonlinearity of the material varies with incident photon energy, consistent with prior studies on second-harmonic generation at different pump wavelengths.[9]

The complex refractive index of bulk NbOI$_2$ and the polarization of the emitted THz radiation was also characterized by THz time-domain spectroscopy (TDS) along polar and non-polar axes. The results show a real part of the refractive index of 2.6 in the range of 1 to 2 THz and absorption features are consistent with our emission results shown in Fig. 2, which also confirms that these modes are IR-active optical phonon. Additionally, the material exhibits significant anisotropy near the absorption peak along the two crystallographic axes. We attribute the ferron oscillations mainly to an impulsive stimulated Raman scattering process under below-bandgap excitation (800 nm excitation)[23]. Meanwhile, the same ferron modes are also observed under above-bandgap excitation (400 nm excitation). In this case, the ferron oscillations originate from the displacive excitation of coherent phonons mechanism[24].

To demonstrate that NbOI$_2$ is an efficient THz emitter, we compare the THz emission from an 8-$\mu$m-thick NbOI$_2$ thin film to that from a conventional semiconductor THz emitter, specifically a 1-mm-thick <110> ZnTe crystal, as shown in Fig. 3(d). First, we optimize azimuthal angles of both samples to maximize the emission efficiency and measured their THz emission as a function of pump fluence under an 800 nm laser. As shown in Fig. 3(d), the ZnTe crystal saturates rapidly (at approximately 4 mJ/cm$^2$) with increasing pump fluence, while no saturation effect was observed for NbOI$_2$ up to 9 mJ/cm$^2$. The inset shows the absolute THz amplitude prior to thickness normalization as a function of pump fluence.

$$\eta_{THz} = \left( \frac{E_{NbOI_2}^{peak}/t_{NbOI_2}}{E_{ZnTe}^{peak}/t_{ZnTe}} \right)^2 \qquad (2)$$

In this measurement, based on the maximum detected amplitude in the inset of Fig. 3(d), the maximum obtained THz electric field is 12 kV/cm of NbOI$_2$, and it is 2.3 times greater than that of ZnTe. However, the thickness is 125 times thinner. By normalizing this maximum THz amplitude to the material thickness, we obtain a maximum ratio of 287. Squaring this ratio to convert amplitude to energy (as shown in the formula above, where $\eta$ is the normalized efficiency, $E$ is the peak electric field, and $t$ is the sample thickness), we find that the maximum energy conversion efficiency per unit thickness of NbOI$_2$ is 8.2×10$^4$ (~10$^5$) times greater than that of a 1-mm-thick ZnTe crystal, as shown in Fig. 3(d). To estimate the ratio of optical rectification efficiency, we selected the ratio of the THz amplitudes in the linear regime, specifically at the lowest pump fluence shown in the inset of Fig. 3(d), then normalized it by the sample thickness, which yields a value of 62.5. Despite this high per-unit-thickness efficiency, its overall efficiency is constrained by the penetration depth of pump light due to material absorption. Our laser power transmission measurements reveal that the penetration depth of the 800 nm laser in NbOI$_2$ is approximately 30 $\mu m$, imposing constraints on scalability. Nevertheless, this penetration depth is three orders of magnitude greater than that of other carrier-based thin-film THz emitters, for example spintronics emitter, transverse thermal effect-based THz emitter, photoconductive antenna.[25,26] Furthermore, previous studies suggest that the penetration depth may increase at longer wavelengths due to reduced absorption[9].

Fig. 3(e) shows the complete THz ferron oscillation as detected by a 500-$\mu m$-thick GaP crystal. A damping time constant of $\tau = 7.5\ ps$ was extracted by fitting the oscillation to a damped oscillator model, which is described by the following equation: $E = A1 * exp\left(\frac{-t}{\tau}\right) * sin(2\omega t)$. These long-lived ferron oscillations were also confirmed by a recent study using an optical pump-probe method.[16] The corresponding FFT spectra are shown in Fig. 3(g) for different pump fluences. We employed quality factor ($Q$-factor, defined as $Q = \omega_0/\Delta\omega$) to quantitatively characterize the ferron oscillations, as it provides a simultaneous measure of both the frequency and lifetime of the ferrons. A maximum $Q$-factor of 110 was obtained at 3.12 THz. With a highly sensitive 1-mm-thick ZnTe detector, we observed oscillations in THz signal with a duration of up to 200 ps (see Fig. 3(f)). The signal at 1.78 THz was stronger, while the response at 3.12 THz was relatively weak, which is attributed to the narrow detection bandwidth of the thicker detector. Under a pump fluence of 2 mJ/cm$^2$, we observed a maximum $Q$-factor of 228, while the peak frequency remained unchanged across different pump fluences. Additionally, while the peak frequency remained constant within our spectral resolution, indicating the robustness of the lattice under high-fluence excitation, the $Q$-factor decreased. This reduction may arise from enhanced scattering between ferrons and excited carriers during relaxation at higher pump fluences[27]. The large-amplitude and long-lived ferron emission observed here is markedly distinct from that in other materials[12,13,28]. It enables the ferron oscillations to dominate the THz emission process, thereby providing enhanced tunability of the THz output through lattice manipulation, such as strain engineering and temperature modulation of the ferronic modes.

**High-efficiency resonant excitation of high-$Q$ ferron through THz high-$Q$ spectroscopy**

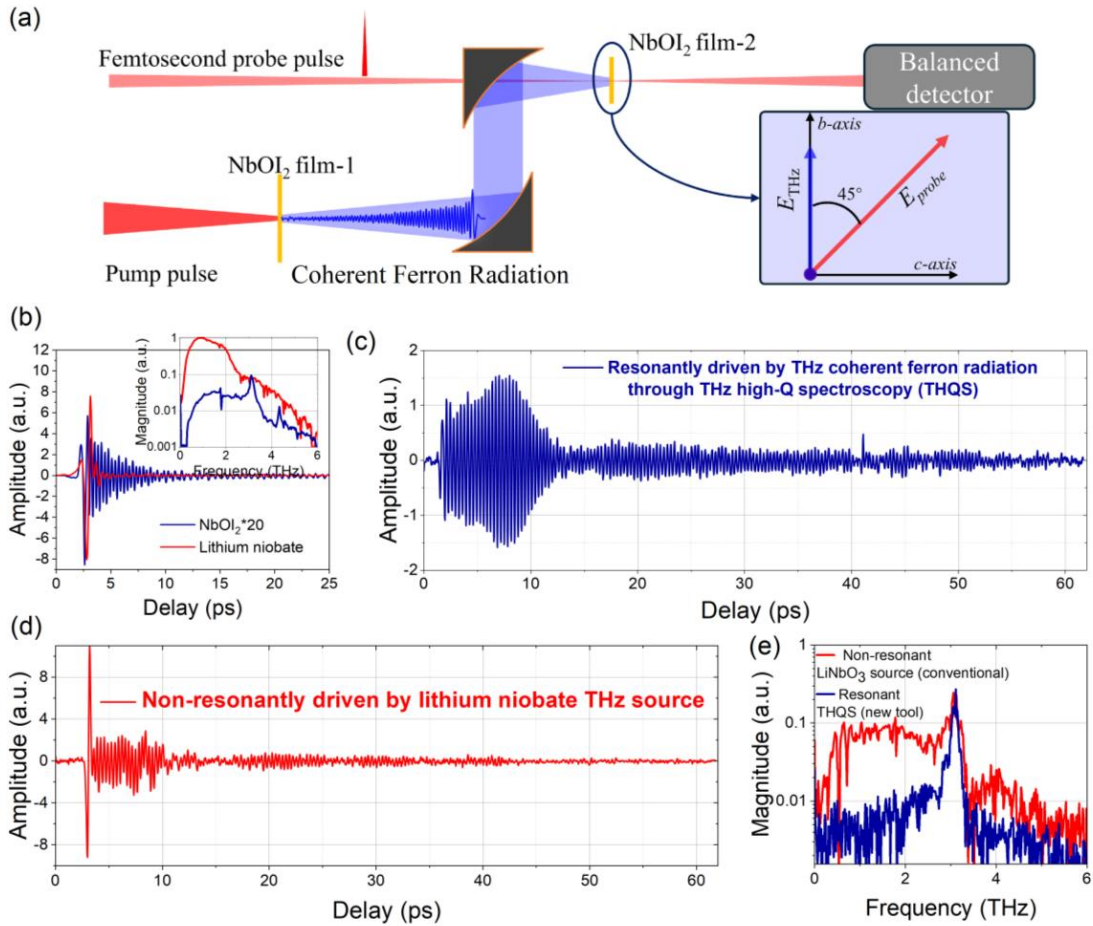

**Fig. 4. High-efficiency resonant excitation of high-$Q$ Ferrons using THz high-$Q$ spectroscopy (THQS). (a)** Experimental layout for resonant excitation of high-$Q$ ferron in NbOI$_2$ film. The NbOI$_2$ film-1 is first pumped by an 800 nm femtosecond laser, generating

narrowband THz coherent radiation which is collected and refocused by a parabolic mirror onto a second NbOI$_2$ film as a resonant Ferronic detector, along with a probe pulse polarized at 45 degrees relative to the THz field (inset). The polarization modulation is monitored by a balanced detector. **(b)** Time domain electric field of the NbOI$_2$ THz source and lithium niobate THz source via tilted pulsefront technique. Amplitude of the NbOI$_2$-based THz source is enlarged 20 times for clear observation and both spectra are shown in the inset. **(c)** Ferron amplitude driven by the narrowband THz coherent radiation from NbOI$_2$ film-1. **(d)** Signal driven by the intense broadband lithium niobate THz source. **(e)** Spectrum of both oscillations in (c) and (d).

The THz ferron radiation generated from these ferroelectric materials is narrowband THz source, which is very suitable for fundamental low-frequency motions excitation including molecular rotation and crystal lattice vibrations, such resonant interaction can permit effective control over matter.[29] Comparing to ultrashort broadband THz source, a narrowband source has the advantage of high spectra density. Here, we demonstrate that THz coherent ferron radiation from one NbOI$_2$ film effectively and resonantly excites the high-$Q$ ferron in a separate NbOI$_2$ film. The experiment layout is shown in fig. 4(a), a femtosecond laser pulse first excites the NbOI$_2$ film-1, generating THz radiation. This emitted THz wave is then collected and refocused onto a second sample, NbOI$_2$ film-2. Simultaneously, a separate femtosecond probe pulse is used to monitor the ultrafast dynamics within the film-2 through electric field induced birefringence effect, which are detected using a balanced photodetector. We compare the ferron excitation effect driven by a conventional lithium niobate (LN) THz source via titled pulse front technique and our narrowband THz ferron radiation source. The corresponding time-domain electric field oscillations and their spectra are shown in Fig. 4(b). The electric field amplitude of LN THz source is 20 times higher than that of THz ferron source and the integration of the power spectrum shows that the energy of LN THz source is 160 times larger. Figures 4(c) and 4(d) illustrate the ferron oscillations driven resonantly and non-resonantly by the two THz sources, respectively. In the resonant case, the ferron vibration amplitude initially builds up under sustained resonant driving, eventually saturating and then decaying due to damping effects. In contrast, in the non-resonant case, the lattice is driven impulsively to a large initial amplitude, after which it undergoes free oscillations, as shown in Fig. 4(d). Even though the energy input of the THz ferron source is two orders of magnitude lower than that of the LN THz source, the vibrational spectra show comparable amplitudes for both sources, as shown in Fig. 4(e). This striking result highlights the high efficiency of resonant excitation enabled by the THz ferron radiation source providing a new THz high-$Q$ spectroscopy (THQS) tool for probing quantum matter.

**Electric-field control of the giant Ferrons**

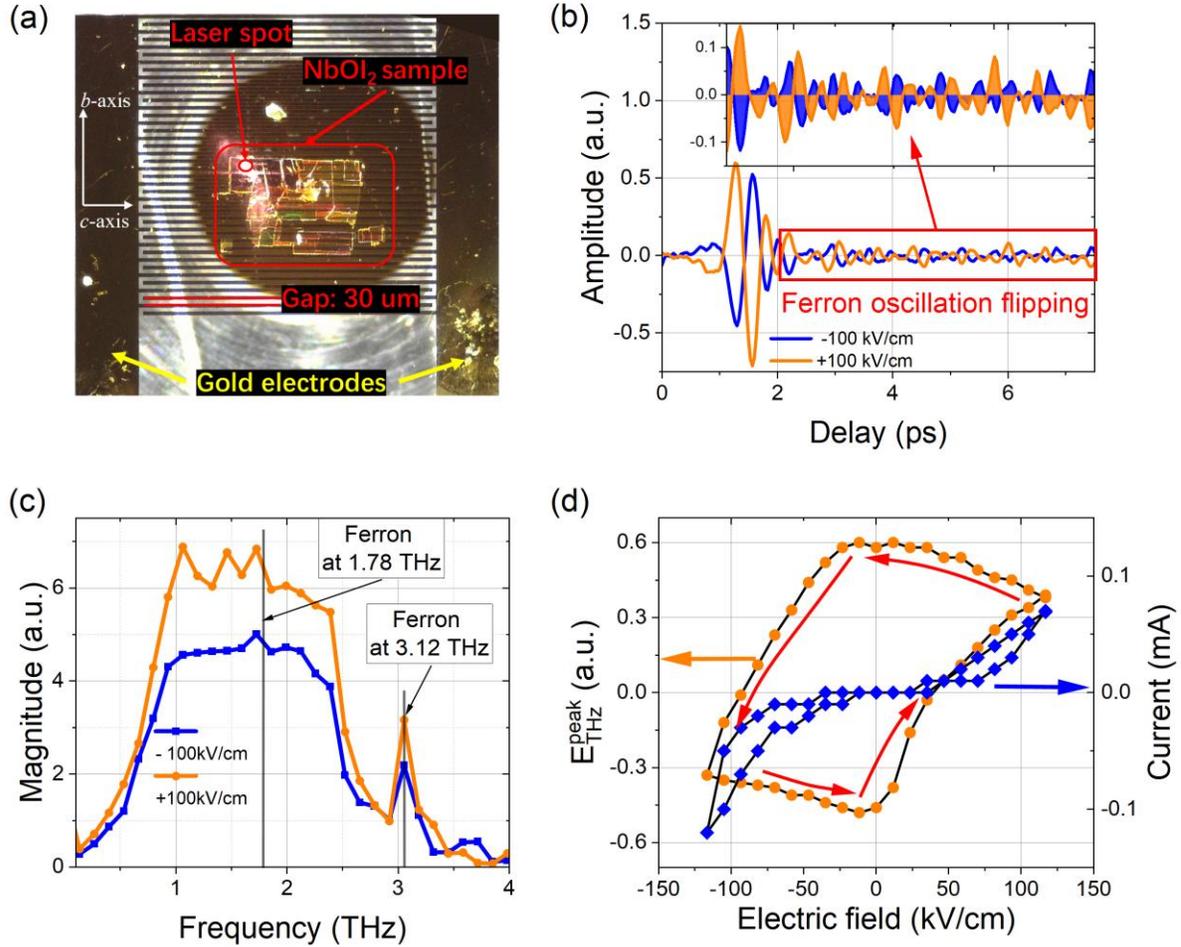

**Fig. 5. Electric-field control of the giant Ferrons. (a)** Experimental configuration for the electric field manipulation of ferrons, photo taken by a microscope. NbOI$_2$ thin film is exfoliated on the interdigital electrode (IDE) with the spontaneous polarization align with the applied electric field, the adjacent electrode gap is 30 $\mu m$. An 800 nm femtosecond laser is incident on one of the gaps and the emitted THz wave is measured by a 1-mm-thick ZnTe crystal. **(b)** Ferron oscillation flipping under an opposite electric field of 100 kV/cm, powered by a high voltage sourcemeter (Keithley 2470, Tektronix). Inset figure show the zoomed area of ferron oscillation for better observation. **(c)** FFT spectrum of **(b)** shows clear ferron mode at 3.12 THz and 1.78 THz. **(d)** Electric hysteresis of emitted THz field and current, red arrows show the sweeping direction of electric field.

Analogous to the magnetic-field control of magnons, direct response to an external electric field provide unambiguous evidence for the existence of ferrons and offer a powerful means to manipulate them.[4] To demonstrate this, we exfoliated NbOI$_2$ thin films onto interdigital electrodes fabricated on a quartz substrate, with an electrode gap of 30 $\mu m$, as shown in Fig. 5a. An 800 nm femtosecond laser was focused onto one of the electrode gaps from the backside, and the emitted THz radiation was detected using a 1-mm-thick ZnTe crystal. By reversing the applied electric field, we observed a clear THz signal in which the ferron oscillations switched polarity (Fig. 5b), and the corresponding spectral features are shown in Fig. 5c where no ferron frequency change is observed. This field-induced switching corresponds to a reversal of the spontaneous polarization, arising from the displacement of Nb atoms in NbOI$_2$, which is consistent with the electrically responsive nature of ferrons. Furthermore, continuous sweeping of the electric field enables non-volatile control of the emitted THz amplitude, producing a pronounced hysteresis loop (Fig. 5d) that reflects the ferroelectric nature of the material. A similar hysteresis was observed in the simultaneously measured current, in agreement with previous reports.[31] Unlike earlier indirect evidence, such as electrically dependent thermal conductivity mediated by ferrons in ferroelectric materials,[4] our direct observation of ferron phase reversal and its associated electrical hysteresis provides definitive proof of their existence.

**Conclusion.**

Our comprehensive study provides direct evidence for the long-theorized Ferrons at THz frequencies. Specifically, we (i) observe coherent THz radiation from the van der Waals ferroelectric NbOX$_2$ (X = I, Br, Cl) and (ii) phase reversal of Ferronic radiation driven by an electric field. The ferronic radiation is dominated by long lived, multiple high-$Q$ ferron modes in NbOX$_2$, and the emission efficiency per unit thickness under 800 nm excitation is five orders of magnitude higher than that of ZnTe, with scope for further enhancement via higher energy optical pump and strain engineering. We identify distinct emission mechanisms, namely optical rectification at 800 nm and ultrafast shift current generation at 400 nm, and link ferron generation to impulsive stimulated Raman scattering and displacive coherent phonon excitation. We further show that NbOI$_2$ exhibits robust temperature stability and strain enhanced THz emission. Using this narrowband THz source to resonantly pump another ferron mode, we achieve an excitation efficiency two orders of magnitude higher than that of high intensity THz source based on lithium niobate. Finally, we demonstrate nonvolatile electric field control of ferron oscillations, further validating the existence and tunability of ferrons. Leveraging the van der Waals ferroelectric, and piezoelectric properties of NbOX$_2$, these class of materials promise electrically tunable, on chip THz sources and resonant control of quantum excitations, laying the groundwork for ferronic polariton THz lasers and next generation integrated THz technologies.

We also note a closely related arXiv preprint by Zhu *et al*. addressing a similar problem.[32] Rather than directly observing polarization-wave propagation, our study provides direct evidence for the existence of ferrons by demonstrating their electric-field manipulation.

**Materials and Methods**
**Materials growth**: NbOX$_2$ (X = I, Br, Cl) crystals are grown by the chemical vapor transport (CVT) method. The stoichiometric ratio (Nb: O: X= 1: 1: 2) of Nb, Nb$_2$O$_5$ with different halogen sources (NbCl$_5$, Br$_2$, I$_2$) are sealed in silica tubes under high vacuum ($10^{-2}$ Pa). The sealed tubes are put in a two-zone furnace, which is heated to 650 °C and 640 °C within 10 hours, and soaked for 72 hours. After 15 days, rectangular crystals (NbOX$_2$) are obtained in the cold zone.

**Raman characterization**: Raman spectra for these exfoliated nanoflakes are carried out on a WITec CRM200 confocal Raman microscopy system with 532 nm excitation laser at room temperature.

**Experimental setup**
A Ti: sapphire femtosecond amplifier laser, operating at a repetition rate of 1 kHz with a pulse width of 45 fs and a maximum output power of 5 W, was employed for the THz ferron emission experiment. The primary laser pulse was divided into two components: one serving as the pump pulse and the other as the probe pulse. The pump pulse was used to excite the NbOX$_2$ sample, while the probe pulse measured the THz electric field via a balanced detection system, which comprised a quarter-wave plate, a Wollaston prism, and a balanced photodetector. The generated THz radiation was collected by a pair of parabolic mirrors and subsequently focused onto an electro-optic crystal (EOC), either 1 mm ZnTe or 500 $\mu m$ GaP, along with the probe laser pulse. For refractive index measurement, we use a 500 $\mu m$ GaP as a THz emitter to obtain a broadband THz source.

**THz Field strength measurement**

The THz electric field strength was measured using the experimental setup described above, based on the linear electro-optic effect. The differential phase retardation $\Delta\phi$ experienced by the probe beam due to this effect over a detector thickness L is given by: $\Delta\phi = \frac{\omega L}{c} n_0^3 r_{41} E_{THz}$, where $\omega$ is the angular frequency of the probe laser, $c$ is the speed of light in vacuum, $n_0$ is the refractive index of the detector at the probe wavelength, $r_{41}$ is the electro-optic coefficient (for GaP crystal, $r_{41} \approx 1\ pm/V$, and $E_{THz}$ is the amplitude of the THz electric field. The differential signal measured by the balanced photo detector is directly proportional to this induced phase retardation:

$$\frac{I_s}{I_0} = \frac{I_a - I_b}{I_a + I_b} \approx \Delta\phi \qquad (4)$$

$$E_{THz} = \frac{I_s}{I_0}\left(\frac{c}{\omega L n_0^3 r_{41} E_{THz}}\right) \qquad (5)$$

Where $I_s$ is the differential signal shown in the lock-in amplifier. $I_0$ is the total signal on each diode of the balanced detector.

**Pump wavelength-dependent experiments**

An optical parametric amplifier (OPA, Light Conversion, covering the 950–580 nm range) was employed to pump the NbOI$_2$ thin film. During wavelength tuning, the movement of internal optical elements within the OPA introduces a timing offset between the pump and probe pulses. To monitor and correct this offset, we use a spintronic THz emitter (CoFe(5 nm)/Pt(3 nm)) as a reference and applied the corresponding time-delay compensation to the NbOI$_2$ THz emission measurements. Since the THz emission from the spintronic emitter is independent of the pump wavelength, it does not exhibit any temporal shift at the fixed pump fluence when the pump wavelength is varied. A 400 nm pump was generated through second-harmonic generation, and a spintronic THz emitter was similarly employed to compensate for the timing offset in this configuration.

**Acknowledgments:** We acknowledge the support from the National Research Foundation (NRF) Singapore, grant no. NRF-CRP23-2019-0005 (TERACOMM) and Grant No: NRF-MSG-2023-0002. Z.L. acknowledges supports from Singapore National Research Foundation–Competitive Research Program NRF-CRP22-2019-0007 and NRF-CRP21-2018-0007. This research is also supported by A*STAR SERC MTC Programmatic Fund (award no. M23M2b0056) and the Singapore Ministry of Education Tier 3 Programmatic Fund (award no. MOE-MOET32023-0003).

**Author contributions:** BL.Z., and R.S. proposed the idea of electric-field control of giant ferronics and THz generation from Niobium Oxydihalide. DR.H. and Z.L. grew the NbOX$_2$ samples. BL.Z. carried out the THz generation measurements with assistance by MS.S., S.J., K.J., TT.C., YJ.T., WH.W. and PT.C.I. DR.H performed the Raman scattering experiment. BL. Z, and

R. S. wrote the manuscript with input from all authors. Z. L. and R.S. co-lead and co-supervised the project.

**Competing interests:** The authors declare that they have no competing interests.

**Data and materials availability:** All the data supporting the findings of this study are openly available in NTU research data repository DR-NTU. Additional information related to this paper is available from the corresponding author, R.S., upon reasonable request.

**References**
1. P, Tang. et al. Excitations of the ferroelectric order. *Phys. Rev. B* **106**, (2022).
2. G, E, W, Bauer. et al. Theory of Transport in Ferroelectric Capacitors. *Phys. Rev. Lett.* **126**, 187603 (2021).
3. A, V, Chumak. et al. Magnon spintronics. *Nature Phys* **11**, 453–461 (2015).
4. B, L, Wooten. et al. Electric field–dependent phonon spectrum and heat conduction in ferroelectrics. *Science Advances* **9**, eadd7194 (2023).
5. E, Rongione. et al. Emission of coherent THz magnons in an antiferromagnetic insulator triggered by ultrafast spin–phonon interactions. *Nat Commun* **14**, 1818 (2023).
6. Q, Guo. et al. Ultrathin quantum light source with van der Waals $NbOCl_2$ crystal. *Nature* **613**, 53–59 (2023).
7. W, Chen. et al. Extraordinary Enhancement of Nonlinear Optical Interaction in $NbOBr_2$ Microcavities. *Adv. Mater.* **36**, 2400858 (2024).
8. I, Abdelwahab. et al. Giant second-harmonic generation in ferroelectric. *Nat. Photonics* **16**, 644–650 (2022).
9. Y, Fang. et al. 2D $NbOI_2$: A Chiral Semiconductor with Highly In-Plane Anisotropic Electrical and Optical Properties. *Adv. Mater.* **33**, 2101505 (2021).
10. Y, Wu. et al. Data-driven discovery of high performance layered van der Waals piezoelectric $NbOI_2$. *Nat. Commun.* **13**, 1884 (2022).
11. W,-C, Chu. et al. Widely linear and non-phase-matched optical-to-terahertz conversion on GaSe:Te crystals. *Opt. Lett.* **37**, 945 (2012).
12. B, Guzelturk. et al. Dynamically Tunable Terahertz Emission Enabled by Anomalous Optical Phonon Responses in Lead Telluride. *ACS Photonics* **8**, 3633–3640 (2021).
13. B, Guzelturk. et al. Terahertz Emission from Hybrid Perovskites Driven by Ultrafast Charge Separation and Strong Electron–Phonon Coupling. *Adv. Mater.* **30**, 1704737 (2018).
14. A, S, Sinko. et al. Polarization sensitive raman scattering and stimulated terahertz emission from GUHP molecular crystal. *IEEE Trans. Terahertz Sci. Technol.* **13**, 526–538 (2023).
15. A, Castellanos-Gomez. et al. Local strain engineering in atomically thin $MoS_2$. *Nano Lett.* **13**, 5361–5366 (2013).
16. C,-Y, Huang. et al. Coupling of electronic transition to ferroelectric order in a 2D semiconductor. *Nat. Commun.* **16**, 1896 (2025).
17. C, Kittel. and P, McEuen. *Introduction to Solid State Physics* (John Wiley & Sons, 2018).
18. M, Tong. et al. Ultraefficient Terahertz Emission Mediated by Shift-Current Photovoltaic Effect in Layered Gallium Telluride. *ACS Nano* **15**, 17565–17572 (2021).
19. B, Mortazavi. et al. Highly anisotropic mechanical and optical properties of 2D $NbOX_2$ (X = Cl, Br, I) revealed by first-principle. *P. Soc. Photo-opt. Ins.* **33**, 275701 (2022).
20. Y, Jia. et al. Niobium oxide dihalides $NbOX_2$: a new family of two-dimensional van der Waals layered materials with intrinsic ferroelectricity and antiferroelectricity. *Nanoscale Horiz.* **4**, 1113–1123 (2019).
21. C, Liu. et al. Ferroelectricity in Niobium Oxide Dihalides $NbOX_2$ (X = Cl, I): A Macroscopic- to Microscopic-Scale Study. *ACS Nano* **17**, 7170–7179 (2023).
22. M, Sotome. et al. Spectral dynamics of shift current in ferroelectric semiconductor SbSI. *Proc. Natl. Acad. Sci.* **116**, 1929–1933 (2019).
23. Y,-X, Yan. et al. Impulsive stimulated scattering: General importance in femtosecond laser pulse interactions with matter, and spectroscopic applications. *J. Chem. Phys.* **83**, 5391–5399 (1985).
24. C, T. K. et al. Mechanism for displacive excitation of coherent phonons in Sb, Bi, Te, and $Ti_2O_3$. *Appl. Phys. Solids Surf.* **55**, 482–488 (1992).
25. T, S, Seifert. et al. Spintronic sources of ultrashort terahertz electromagnetic pulses. *Appl. Phys. Lett.* **120**, 180401 (2022).

26. W, Lu. et al. Ultrafast photothermoelectric effect in Dirac semimetallic $Cd_3As_2$ revealed by terahertz emission. *Nat. Commun.* **13**, 1623 (2022).
27. Y, Han. et al. Photoinduced Ultrafast Symmetry Switch in SnSe. *J. Phys. Chem. Lett.* **13**, 442–448 (2022).
28. J, Afalla. et al. Terahertz emission from transient currents and coherent phonons in layered $MoSe_2$ and $WSe_2$. *J. Appl. Phys.* **133**, 165103 (2023).
29. T, Kampfrath. et al. Resonant and nonresonant control over matter and light by intense terahertz transients. *Nat. Photonics* **7**, 680–690 (2013).
30. N, A, Lanzillo. et al. Temperature-dependent phonon shifts in monolayer $MoS_2$. *Appl. Phys. Lett.* **103**, 093102 (2013).
31. Liu, Q. et al. Lowering the Coercive Field of van Der Waals Ferroelectric NbOI2 with Photoexcitation. *Appl. Phys. Lett.*, *126*, 43104 (2025)
32. Choe, J. et al. Observation of coherent ferrons. Preprint at https://doi.org/10.48550/arXiv.2505.22559 (2025).